\documentclass{cimento}

\usepackage{graphicx}  

\title{Roberto Petronzio and the QCD}
\author{Giorgio Parisi\from{ins:x}}
\instlist{\inst{ins:x} 
  Dipartimento di Fisica, Universit\`a di Roma - Roma, Italy}

\begin{document}

\maketitle

\begin{abstract}
This paper aims to recall some of the main contributions of Roberto Petronzio to QCD, with a particular regard to the period we have been working together. His seminal contributions span both the development of analytic computations using perturbation theory and the starting of lattice gauge theories.
\end{abstract}

\hfill {\sl Why should we work on this problem if we do not have fun?}

\hfill Nicola Cabibbo
\section{First papers}

The fist paper of Roberto was {\sl The nucleon as a bound state of three quarks and deep inelastic phenomena}. It appeared in August 1973 \cite{ACMP74}. 

It was based on the very nice idea of describing the quarks wave function inside the nucleon in the $p=\infty$ frame using information coming from internal symmetries like $SU(6)$. The paper was later extended \cite{ACMP74A,ACMP75} to get predictions for other processes like neutrino scattering and lepton production in proton-proton collisions.

It was a very interesting  paper: \begin{itemize}
\item Good models for the parton distribution were quite rare at that time. The paper describes may be the first reasonable model valid not only for the quarks but for also for the {gluonic structure function}. Gluonic structure functions will be later crucial for computing scaling violations via the process of fragmentations into quark.
\item The model incorporates the knowledge that that time people had on symmetries, not only $SU(3)$, but also  $SU(6)_W$.
\item It stresses the importance of the  $p=\infty$ frame, that will play a very important role in understanding scaling violations in a parton model framework in later years.
\end{itemize}

The paper assumed that the physical octet of Barions was the combination of octets belonging to the following representations of $SU(6)$, the $56,l=0$ and the $70,l=1$. In the case of a nucleon with spin component  $J_z=1/2$, this Barion should be a linear combination of the following three states:
\begin{eqnletter}
 \label{e.all}
     |A\rangle&=& |8,\frac12,\frac12,0,0\rangle_{56}\\
 |B\rangle&=&-\frac{1}{\sqrt{3}}|8,\frac12,\frac12,1,0\rangle_{70}
 -\sqrt{ \frac23}|8,\frac12,-\frac12,1,1\rangle_{70}\\
 |C\rangle&=&\frac{1}{\sqrt{2}}|8,\frac32,\frac32,1,-1\rangle_{70}+
 \frac{1}{\sqrt{3}}|8,\frac32,\frac12,1,0\rangle_{70}+
 \frac{1}{\sqrt{6}}|8,\frac32,-\frac12,1,1\rangle_{70}
\end{eqnletter}
If only the 56 representation were present, one would obtain the bound
\begin{equation}
\frac32\ge\frac{F_2^{eN}}{F_2^{eP}}\ge\frac23 \,,
\end{equation}
that is violated by the experimental data, hence the need of introducing the mixing with the 70 representation.

Our {first paper together} \cite{ACMPP74} was quite unfortunate: {\sl Is the 3104 MeV vector meson the $\psi_c$ or the $W_0$?} It was signed by {G. Altarelli, N. Cabibbo, R. Petronzio, L. Maiani, G. Parisi.}
The paper presented a nice phenomenological analysis: at the end of the paper, we concluded that the 3104 MeV vector meson was the  $W_0$, an answer that is factually wrong, in spite of the elegant arguments in the paper.

I think that we should not be ashamed for writing such a deadly wrong paper. In this I am comforted by the judgment  of Shelly Glashow:
{\sl It is the business of theorists to speculate, and we often find that our speculations are wrong. I have published more than a few papers that have turned out to have been wrong. So have most of my colleagues. That's the name of the game! }(...) 
{\sl Scientists publish speculative results not because they are true, but because they may be true. If they refrained from publishing their speculations for fear that they may not always be true, there would be little progress in science. Even our greatest heroes, Galileo, Newton and Einstein, have published speculations that turned out to be quite false.}

\begin{figure}[!t]
  \centering
  \includegraphics[width=.6\columnwidth]{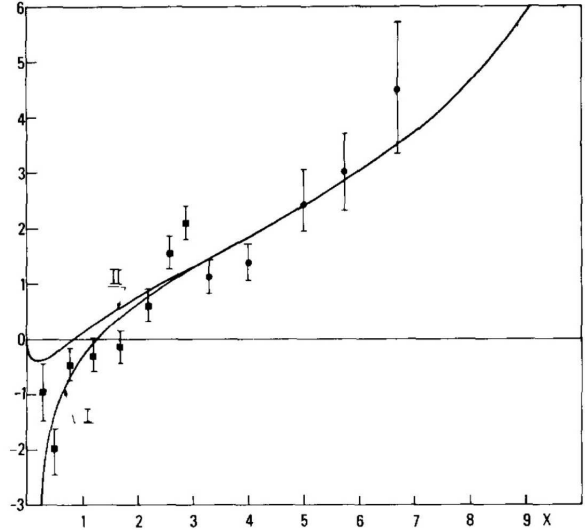}
 \caption{Curve I is our prediction for $d \ln F^p_x(x,q^2)/d \ln q^2$ compared with the experimental data. Curve II is obtained retaining only the octect operators in the operator expansion (taken from \cite{PP76}).
 }
 \label{fig:local}
\end{figure}
 
 Our collaboration went on producing more interesting results. Maybe our best paper of the Roman period was {\sl On the breaking of Bjorken scaling} \cite{PP76}. This paper contains the first computation of scaling violations in QCD taking care of the presence of Gluons. The paper was built on Roberto's great experience on parton wave functions inside the nucleon, especially on the gluonic contribution that was an essential component for having an agreement with the experimental results at small $x$: in this region gluon fragmentation is the dominant process.  It is remarkable that the computation was done 1976, before the AP (Altarelli Parisi) evolution equations \cite{AP77} \footnote{This paper proves that one can compute scaling violations without using the AP equations.}. 
 
 Roberto Petronzio continued to work on the problem of scaling violations in deep inelastic scattering. Two years later he wrote with Nicola Cabibbo {\sl The Two-stage model of hadron structure: Parton distributions and their $Q^2$ dependence} \cite{CP78}, where a similar a more accurate analysis was done, now using the AP equations.

 \section{At CERN}
 Roberto went to CERN in 1977. Most of his works of the first years in CERN were on QCD and weak interactions.
 
 The theoretical panorama on QCD had changed. The AP equations have separated the study of quarks from the rather complex light cone expansion \cite{BP}: in these equations we find  the effective parton distribution that  should be universal and they should be the same in all processes. However finite perturbative corrections proportional to the running coupling constant $\alpha(q^2)$ were supposed to present and to be process-dependent.
 
This picture was easy to conjecture, but it was not easy to prove. The proof finally came in a seminal work with deep theoretical consequences:  {\sl Relating hard QCD processes through the universality of mass singularities} \cite{CP78} by Amati, Petronzio and Veneziano. Here I will not describe this very important paper because it is amply discussed in the contribution of Gabriele Veneziano in this issue.
  
 The point-like nature of QCD implied the existence of jets, of power (fat) tails in the transverse momentum distributions. However, at the time of that paper, the energy of the colliding particles was not high enough to see in a clear way the jets in the final states, also because the quark energy is partitioned between many hadrons via the process of jet fragmentation and quark recombination. On the other hand in the so called Drell-Yan  process, i.e.
 \begin{equation}
p+p \to \mbox{hadrons}+\mu^++\mu^-\,,
\end{equation}
the transverse momentum of the $\mu^++\mu^-$ pair is the same of the quark antiquarks pair that produces a virtual photon: this process allows us to measure the transverse momentum spread or the quarks inside the proton.
 These considerations explain why first hard scattering QCD contributions were computed  for the Drell-Yan process: the prediction was quite neat without having to discuss the quark fragmentation process. 
 
 Two crucial seminal contributions were given by Roberto  in 1978 with the papers  
 {\sl Transverse momentum of muon pairs produced in hadronic collisions} \cite{APP78a} and {\sl Transverse momentum in Drell-Yan processes} \cite{APP78b} written by Altarelli, myself and Roberto. A careful job was done in studying the increase of the average  transverse momentum squared ($p_T^2$)  as a function of $Q^2$ and of the various physical parameters.
 A problem that we had to face was the separation of the two contributions: the one coming from intrinsic spread of the quark wave function inside the nucleon and the one coming from hard processes.
 \begin{figure}[!t]
  \centering
  \includegraphics[width=.6\columnwidth]{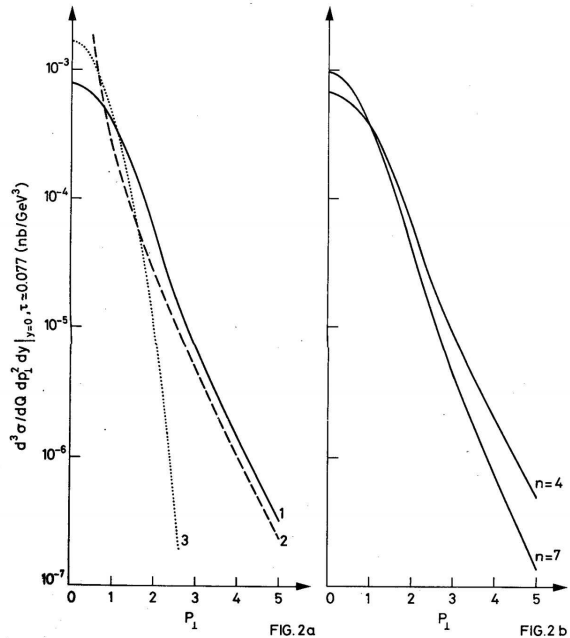}
 \caption{The differential cross section for the Drell-Yan process as function of $p_T^2$ (taken from \cite{APP78b}). Fig. 2a:  curve (1) is our prediction, curve (2) is the one loop contribution and curve (3) is the intrinsic contribution.  Fig. 2b shows how the result depends on the details of the gluon distribution, parametrized by $n$.
 }
 \label{fig:local}
\end{figure}

 Roberto Petronzio was very interested in the resummations of leading logs in special processes, a problem that was studied in the case of QED, but not for QCD. The first paper on this subject is {\sl Heavy flavor multiplicities at very high energies }by Furmanski,  Petronzio and Pokorski \cite{FPP79}. New techniques had to be invented in order to circumvent new problems.
 
 They found the surprising result that the multiplicities increase faster than any power of the log of the energy scale, 
\begin{equation}
\langle n\rangle \propto \exp\left( \sqrt{\alpha_G \log(Q^2/Q_0^2)}\right)
\end{equation}
This result was found quite puzzling by the author themselves, and this  reaction is natural: at that time a simple logarithmic increase of multiplicities was supposed to be established. Nowadays that we know experimentally that the multiplicities increase much faster than a logarithmic of the energy, the results is much less puzzling.
 
A paper that had a long influence was
 {\sl Small transverse momentum distributions in hard processes} by Roberto and myself \cite{PP79}.  The problem was the find the {\sl small} transverse momentum behavior of the  distribution of hard produced muon pairs as an effect of multiple gluon production. In the computation done in \cite{APP78a,APP78b} an intrinsic momentum distribution was needed to avoid the singularity at $p_T=0$ of the first order in perturbation theory. It was clear from the physical viewpoint that multiple gluon production should produce a regularization effect at small momentum, however, the consequences of this phenomenon were not clear.
 
Many ingredients entered in the cocktail  \cite{PP79}. 
\begin{itemize}
\item The leading logs approximation for multiple soft gluon bremsstrahlung.
\item The exponential damping of the elastic form factors.
\item The different behavior of the cross sections in momentum and in impact parameter space.
\end{itemize}
 One of the conclusions of that paper (that I still find surprising) is that the peak at $p_T=0$ flattens with a width proportional to $\left(Q^2\right)^{\gamma}$ with $\gamma=\frac{16}{25}\ln(66/41)\approx 0.305$. The presence of a simple non-integer power of $Q^2$ is quite astonishing in a world dominated by logarithmic corrections to integer powers.
 
 Another paper that had a long and quite likely  larger influence was
 {\sl Singlet parton densities beyond leading order} \cite{FP80} by   Furmanski, and Petronzio.
 
 This was the {\sl manifesto} for next to the leading order computation in QCD. The difficulties tacked in this paper were not only in doing the detailed computations, that were highly nontrivial, but in proving for the first time that those computations were possible.
 The technical tool that they invented was based on the explicit study of the factorization properties of mass singularities. In this way {\sl "within our scheme the predictions for a particular process are obtained by convoluting a universal parton density with a short-distance cross section specific to the process."}
 
 It was the triumph for the marriage of the parton model with QCD.

 These results were extended by to the nonsinglet cases in the papers 
 {\sl Evolution of parton densities beyond leading order: The non-singlet case} \cite{CFP80} by   Curci, Furmanski and  Petronzio  and 
 {\sl Lepton-hadron processes beyond leading order in quantum chromodynamics} \cite{FP82}  by Furmanski and Petronzio.
  \begin{figure}[!t]
  \centering
  \includegraphics[width=.6\columnwidth]{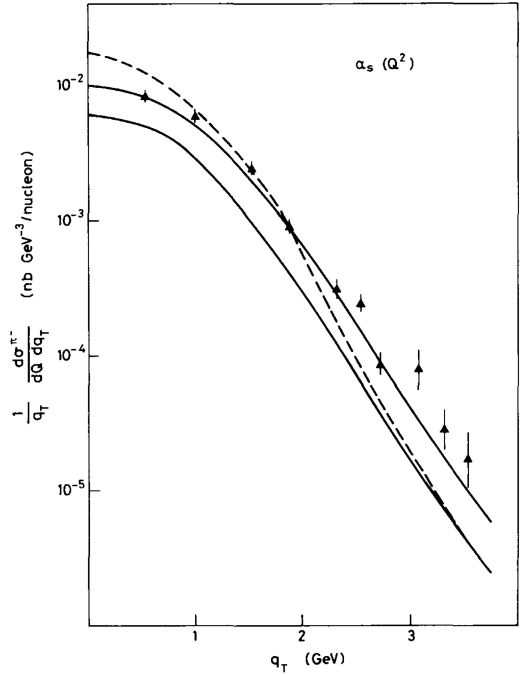}
 \caption{ Experimental data of $1/(q_T)d\sigma^{\pi-N}/(dQ\,dq_T)$ versus $q_T$ at $S=282\, \mbox{GeV}^2$, $Q^2=52.2\,\mbox{GeV}^2$ are compared with the theoretical predictions. The upper full line represents the next-to-leading estimate while the lowest full line represents the lower order estimate (taken from \cite{EMP83}).
 }
 \label{EMP}
\end{figure}
 The  techniques introduced in these papers allowed the computation of next to the leading order results in other processes, as it was done in  the paper by 
 {\sl Lepton pair production at large transverse momentum in second order QCD}  by Ellis,  Martinelli and Petronzio \cite{EMP83}. This is a very remarkable paper, because it contains the first evaluation of next to the leading order effects in QCD for the $p_T$ distribution in the Drell-Yan process. The computation was quite involved because the authors had to compute the $\alpha_s^2$ corrections to a process of order $\alpha_s$. You can see from fig.  (\ref{EMP}) the importance of adding next to the leading effects in order to reach a good agreement with the experimental data.

 Roberto continued to gave seminal contribution to QCD with {\sl Power corrections to the parton model in QCD} \cite{EFP82} and
 {\sl Unravelling higher twists} \cite{EFP83} and  both written with Ellis,  Furmanski (1983).
 
 A paper that was quite ahead of its time was
 {\sl Momentum distribution of $J/\Psi$ in the presence of a quark-gluon plasma} by Karsch and Petronzio \cite{KP87}. The subject of the paper is quite different from the previous ones.  In heavy nuclei collisions at very high energy we could have the formation of a new phase of matter, i.e. {\sl quark-gluon plasma}: there were many theoretical arguments that pointed in that directions. However, it was not clear which was a good experimental signature for this phenomenon. In this paper the author discuss the very interesting suggestion that the momentum distribution of the produced $J/\Psi$ particles should be strongly affected by the phase transition to this new state of matter.
 \section{Lattice QCD}
 
 At the beginning of the eighties, the central interests of Roberto already started to move toward lattice theories and lattice QCD. 
 
The subject was completely new and there was the need of understanding which were the possible artifacts of lattice computations.  The simplest case was the two-dimensional $O(3)$ spin model. The physics of the model was very clear (a ferromagnetic transition that was avoided as an effect of the impossibility of having a Goldstone mode in two dimensions). Moreover, the theory was asymptotically free (like QCD) and topological effects, like instantons, were present in also in this case.  The simplicity of the theory allowed many detailed computations.

 \begin{figure}[!t]
  \centering
  \includegraphics[width=.6\columnwidth]{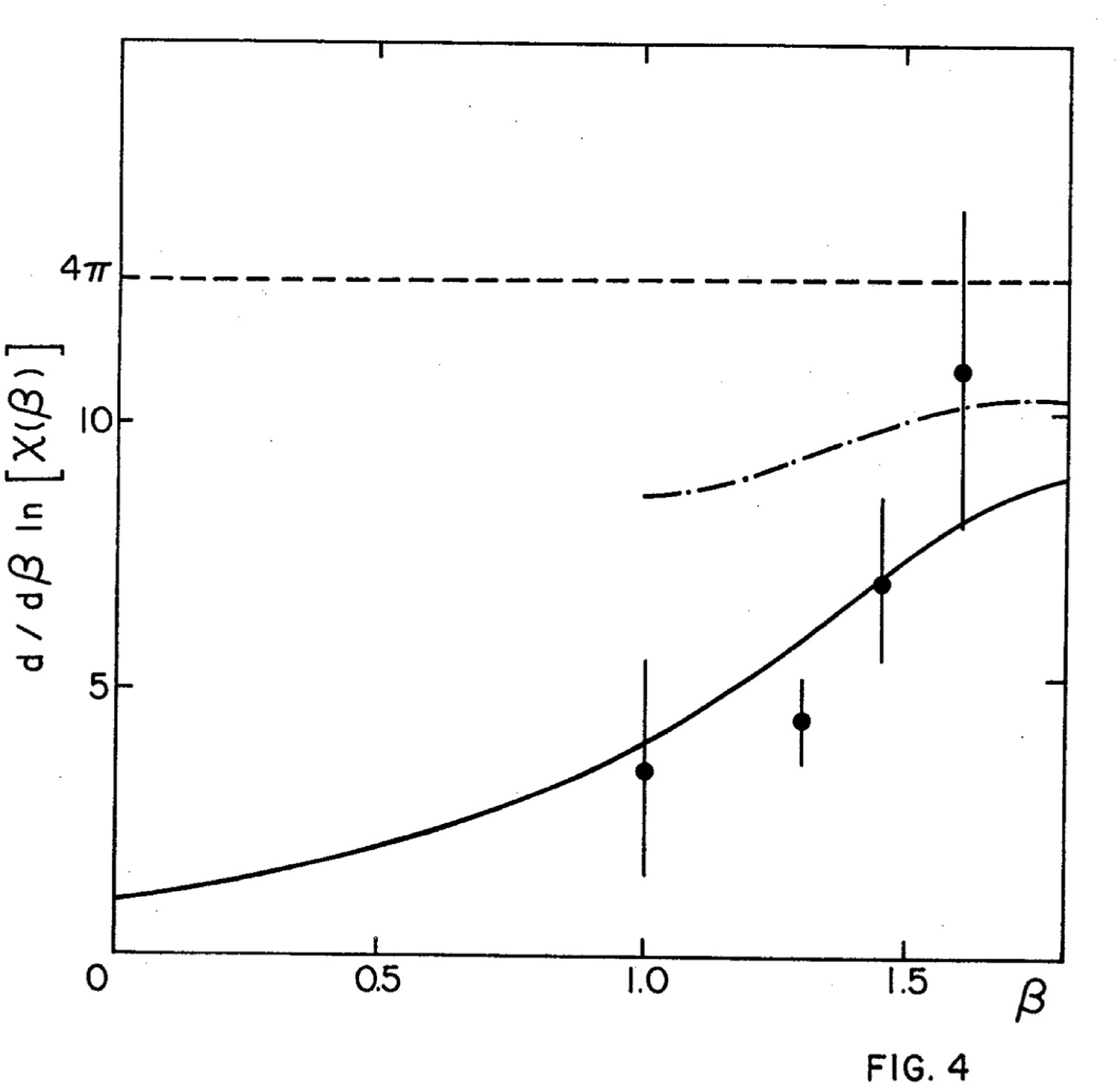}
 \caption{ The quantity $d\log(\chi(\beta))/d\beta$ as function of $\beta$. The dashed line is the asymptotic value at $\beta=\infty$, an the dashed dotted curve takes care of the preasymptotic corrections coming from the next to the leading order (taken from \cite{MPP81}).
 }
 \label{fig:local}
\end{figure}

 Roberto, Martinelli and I started to do Montecarlo simulations for this theory. In our first paper {\sl Monte Carlo simulations for the two-dimensional $O(3)$ nonlinear sigma model} \cite{MPP81}, we tried to study for the first time \footnote{In that years most of the computations were done for the first time, Montecarlo for lattice theories was so a new approach.} the behavior of the magnetic susceptibility $\chi(\beta)$ at high $\beta$ (low temperature). 
 We knew that for large $\beta$ 
 \begin{equation}
\chi(\beta) \propto \beta^{-4} \qquad  \lim_{\beta\to\infty}{d\log(\chi(\beta))\over d\beta}=4\pi
\end{equation}
 We wanted to understand how fast the limit was reached and we were not happy because the approach was quite slow. We used lattices with $L^2$ points with $L$ in the range from 30 to 80. It was clear that we needed a much larger lattice in order to be near to the asymptotic limit. If the same phenomenon were present for QCD, the whole field would be destroyed because four-dimensional lattices with $L=80$ are at the boundary of present day technology.
 
 In order to decrease lattice effects, we found for this model the form of the improved lattice action  where $O(a^2)$ corrections were absent \footnote{As usual $a$ is the lattice spacing.}:  this was done in {\sl Improving the lattice action near the continuum limit} \cite{MPP82}. This computation was done taking care also of one loop corrections that had to be evaluate for the lattice theory. In some sense, we  computed the difference between the one loop results in the continuum and one loop results in the lattice: at the end of the day, we  added counter-terms in order to compensate for the difference of these two computations. This was the first of a huge family of improved actions that have been widely used in QCD and in weak interactions on the lattice.
  
 An other remarkable paper of that time was {\sl Topological charge on the lattice: The $O (3)$ case} \cite{MPV82}, by Martinelli, Petronzio and Virasoro.  In this paper, the authors presented the first definition {the topological charge} for the two-dimensional $O(3)$ spin model, constructed in such a way of not taking care of small instantons. The instanton density was also computed and compared with the numerical results.

 However, most of the fun was with lattice QCD.
 It was a new world that we started to explore with excitement. All the low-energy strong interaction parameters were computable. This was a complete change from the previous situation where only phenomenological arguments can be used, mostly in hand waving arguments. Of course we knew that the measurements were affected by strong systematic effects (we started our computation with a $5^3\times 10$ lattice), however, it was rather surprising to see that all quantities, one after the other, were in qualitative agreement with the experimental data.
 
 There are so many papers in that period that I will just briefly recall them. The collaboration was floating and the author list often changed.
 \begin{itemize}
 \item We started with the computation of the basic properties of hadrons in the quenched approximation in {\sl Hadron spectroscopy in lattice QCD} \cite{FMOPPR82}, where the statistic and systematic errors were strongly reduced with respect to the previous papers.
 \item We computed the {\sl  proton and neutron magnetic moments in lattice QCD} \cite{MPPR82} by measuring the mass splitting in presence of a magnetic field: in this case we found for the gyromagnetic factor of the proton $g_P=3.0\pm 0.6$ versus an experimental value of 2.79 and for the ratio of the gyromagnetic factors of the proton and of the neutron $g_P/g_N=-1.60\pm 0.15$ versus an experimental value of $-1.46$.
 \item We computed  the strange hadron masses \cite{MOPP}, in particular the {\sl $\Lambda-\Sigma_0$ splitting}. Here the result was not too satisfactory: the sign was the correct one, but its absolute value was quite small. We argued that this was an example of a general phenomenon: all the mass splitting due to spin-spin interactions were quite small. We obtained a reasonable value to the ratio
 \begin{equation}
{m_\Sigma-m_\Lambda \over m_\Delta - m_P}=0.18\pm 0.09\,,
\end{equation}
to be compared to the experimental value of 0.26.
\item  In {\sl Boundary effects and hadron masses in lattice QCD} \cite{MPPR83} we identified a relevant contribution to the large fluctuations of hadron masses present in lattice calculations with periodic boundary conditions The contribution is due to unphysical quark paths which are absent in the infinite volume limit. We showed that these contributions can be eliminated by averaging over possible rotations of the boundary links by the elements of the $Z(3)$ subgroup. In this way, the  "effective" volume for these paths is triplicated.
 \end{itemize}

A very remarkable paper was {\sl Hadron spectrum in quenched QCD on a $10^3\times 20$ lattice} \cite{LMPR83} by Lipps, Martinelli, Petronzio and Rapuano.  It was real progress respect to the previous analysis on smaller lattices ($5^3$, $6^3$, $8^3$) and allowed us for the first time to investigate the systematic effect due to non-zero lattice sides.  A subsequent paper was {\sl Kogut-Susskind and Wilson fermions in the quenched approximation: A Monte Carlo simulation} \cite{BMP} where Billoire, Marinari and Petronzio presented a systematic comparison of the results for both the Kogut-Susskind and Wilson fermions.

Roberto was also interested to analyze the behavior of QCD without Fermions. Here the most relevant observable (beyond the glueball mass) is the string tension. However, his precise determination was quite difficult due to large statistical errors. We  computed the string tension \cite{PPR83} with good accuracy. The computation was possible due to a clever trick for noise reduction (multihit) that we introduced and that became a standard tool. The gain induced by the trick was  a decrease of factor 10 in the statistical error, corresponding  to a gain of a factor 100 in time. Roberto continued to work on pure gauge QCD. He wrote with Karsch a very nice paper {\sl Gluon thermodynamics near the continuum limit} \cite{KP} on the quark liberation phase transition (that correspond the formation a quark-gluon plasma), a problem that we have already seen he has analyzed in a later paper with the same author \cite{KP87}.

Roberto was also among the proponents of the first  APE project. He contributed to the first two papers describing the computer {\sl The APE project: a computer for lattice QCD} \cite{APE84} and {\sl The APE project: a gigaflop parallel processor for lattice calculations} \cite{APE85}. Unfortunately, the collaboration with Roberto inside the APE could not  continue due to logistic problems. During the construction of the machine   the work was concentrated in Bologna (memory card), Pisa (controller and local network), Rome (floating point unit and software), but not in CERN. He was strongly involved in using the subsequent APE machines, but this part of the story can be found in the contribution of Martin L\"uscher (this issue).

In the meanwhile a new investigation subject was open by Cabibbo, Martinelli, Petronzio with the paper {\sl Weak interactions on the lattice} \cite{CMP84} where they showed  that lattice QCD can be used to evaluate the matrix elements of four-fermion operators which are relevant for weak decays. This was the starting point of so many computations of weak matrix elements that are very important in the testing of the standard model and in the eventual discovery of new physics. A more detailed analysis of this and other papers by Roberto can be found in the contribution of Luciano Maiani (this issue).

\end{document}